\documentclass[twocolumn,showpacs,preprintnumbers,amsmath,amssymb,floatfix,
superscriptaddress]{revtex4}

\usepackage{graphicx}
\usepackage{graphics}
\usepackage{dcolumn}
\usepackage{bm}
\usepackage{amsbsy} 
\usepackage{SIunits}
\usepackage{setspace}
\usepackage[a4paper]{geometry}

\usepackage{amsmath}
\usepackage{amsfonts}
\usepackage{amsbsy}
\usepackage{amssymb}

\usepackage{bm}
\usepackage{graphicx}

\newcommand{\colvec}[1]{\ensuremath{\mathrm{#1}}}

\newcommand{\spvec}[1]{\ensuremath{\vec{#1}}} 

\begin{document}

\title{Coherent control of nanoscale light localization in metamaterial: creating and positioning a sub-wavelength energy hot-spot}

\author{T. S. Kao}
\affiliation{Optoelectronics Research Centre and Centre for Photonic Metamaterials, University of Southampton, Southampton SO17 1BJ, United Kingdom}
\author{S. D. Jenkins}
\affiliation{School of Mathematics and Centre for Photonic Metamaterials, University of Southampton, Southampton SO17 1BJ, United Kingdom}
\author{J. Ruostekoski}
\affiliation{School of Mathematics and Centre for Photonic Metamaterials, University of Southampton, Southampton SO17 1BJ, United Kingdom}
\author{N. I. Zheludev}
\email{n.i.zheludev@soton.ac.uk}
\affiliation{Optoelectronics Research Centre and Centre for Photonic Metamaterials, University of Southampton, Southampton SO17 1BJ, United Kingdom}

\date{\today}

\begin{abstract}

Precise control and manipulation of optical fields on a nanoscale is one of the most important and challenging problems in ``nanophotonics''. Since optical wavelength is on a much larger microscale, it is impossible to employ conventional focusing for that purpose. We show the strong optically-induced interactions between discrete meta-molecules in a metamaterial system and coherent monochromatic continuous light beam with a spatially-tailored phase profile can be used to prepare a sub-wavelength scale energy localization. Well isolated energy hot-spots as small as $\lambda/10$ can be created and positioned at will on the metamaterial landscape offering new opportunities for data storage and imaging applications.

\end{abstract}
\maketitle


\begin{figure}
  \linespread{1}
  \includegraphics[width=80mm]{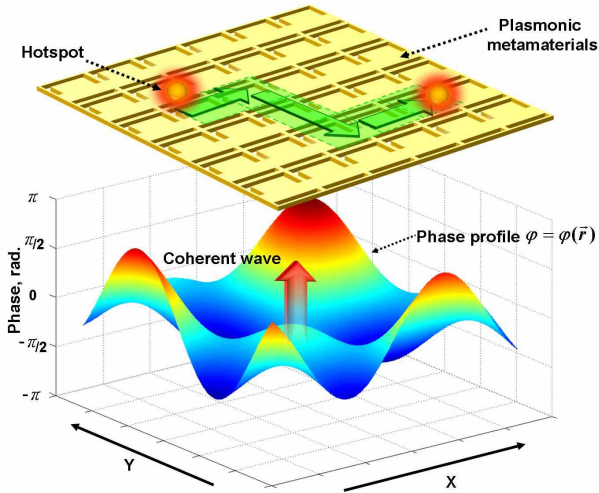}
  \caption{Coherent control of light localization in array of interacting meta-molecules: artistic impression. A light beam with modulated phase profile excites a metamaterial array of interacting plasmonic resonators leading to the formation of the energy hot-spot at the metamaterial. This spot may be moved from a metamaterial cell to another metamaterial cell by tailoring the phase profile of the incident beam.}\label{conf}
\end{figure}

In 2002 a method was suggested, which was based on tailoring phase modulation of ultrashort optical pulses in the time domain to achieve coherent control of spatial distribution of the excitation energy in complex inhomogeneous nanosystems \cite{stockman2002}. In such systems the localized plasmon frequencies vary from nanoscale feature to nanoscale feature and thus can be correlated with their positions. The pulse phase modulation will cause the exciting field to take energy away from surface plasmons localized in those parts of the system where the oscillations are out of phase with the driving pulse and move it, with time, to the surface plasmon excitations in other parts where such oscillations occur in phase with the driving pulse. Later a similar idea based on tailoring of ultrafast pulses in the vicinity of silver nanostructures through adaptive polarization shaping has been successfully demonstrated \cite{schnell2009, aeschlimann2010}. Illuminating a sub-wavelength diffraction grating by an amplitude and phase modulated continuous-wave light beam has been suggested as a method to generate sub-diffraction light spots \cite{SentenacPRL2008}.

Here we show that strong interactions between regularly spaced plasmonic resonators in a planar metamaterial, a two-dimensional array of nanoscale meta-molecules, can result in a sub-wavelength localization of optical energy. A desired nanoscale light hot-spot can be engineered simply by adjusting the far-field spatial profile of an incoming monochromatic coherent continuous light beam. We show by using a simple model
how the localization results from a {\em co-operative} response of the metamaterial array of altimetrically split ring plasmonic resonators, providing physical insight into the energy localization effect. The essential requirement in the sub-wavelength localization process of the incoming wave is that individual meta-molecules act as closely-spaced {\em discrete} scatterers whose strong inter-meta-molecular interactions modify the energy eigenspectrum of the system. This is in contrast with the homogenization approach to electromagnetic properties of metamaterials treated as a continuous medium instead of a collection of discrete meta-molecules.

We show that evanescent field energy can be localized almost solely at a single meta-molecule while a group of neighboring meta-molecules remain unexcited. Now by simply adjusting the spatial phase profile of the incident beam the nanoscale hot-spot can be moved at will from one meta-molecule to another, providing an efficient technique for a sub-wavelength scale optical control and manipulation of a metamaterial system. In contrast with localization techniques relying on laser-pulse excitation \cite{stockman2002, schnell2009, aeschlimann2010} our method is based on a continuous wave source and does not depend on a transient redistribution of energy between nano-objects. This makes the localization in a metamaterial system much simpler to implement. Moreover, it does not require the nanoscale system to be spatially inhomogeneous (such as a rough surface) and works with periodic, regular planar array of identical nano-objects opening the opportunities for \emph{microscopy and data storage applications} which will be discussed below.

The idea of coupled resonators illuminated by a phase-modulated beam has a clear mechanical analogy in two coupled identical oscillators (modeling nano-objects) that may be driven to vastly different amplitudes by setting up a phase delay between otherwise identical coherent mechanical forces driving them. A system consisting of resonant wires has previously been used as a metalens, transferring sub-wavelength features of a evanescent
field to the spectral properties of propagating microwaves at different frequencies that are detectable in the far-field \cite{LemoultPRL10} (see also Ref.~\cite{Tsang07}).

\begin{figure*}
  \linespread{1}
  \includegraphics[width=150mm]{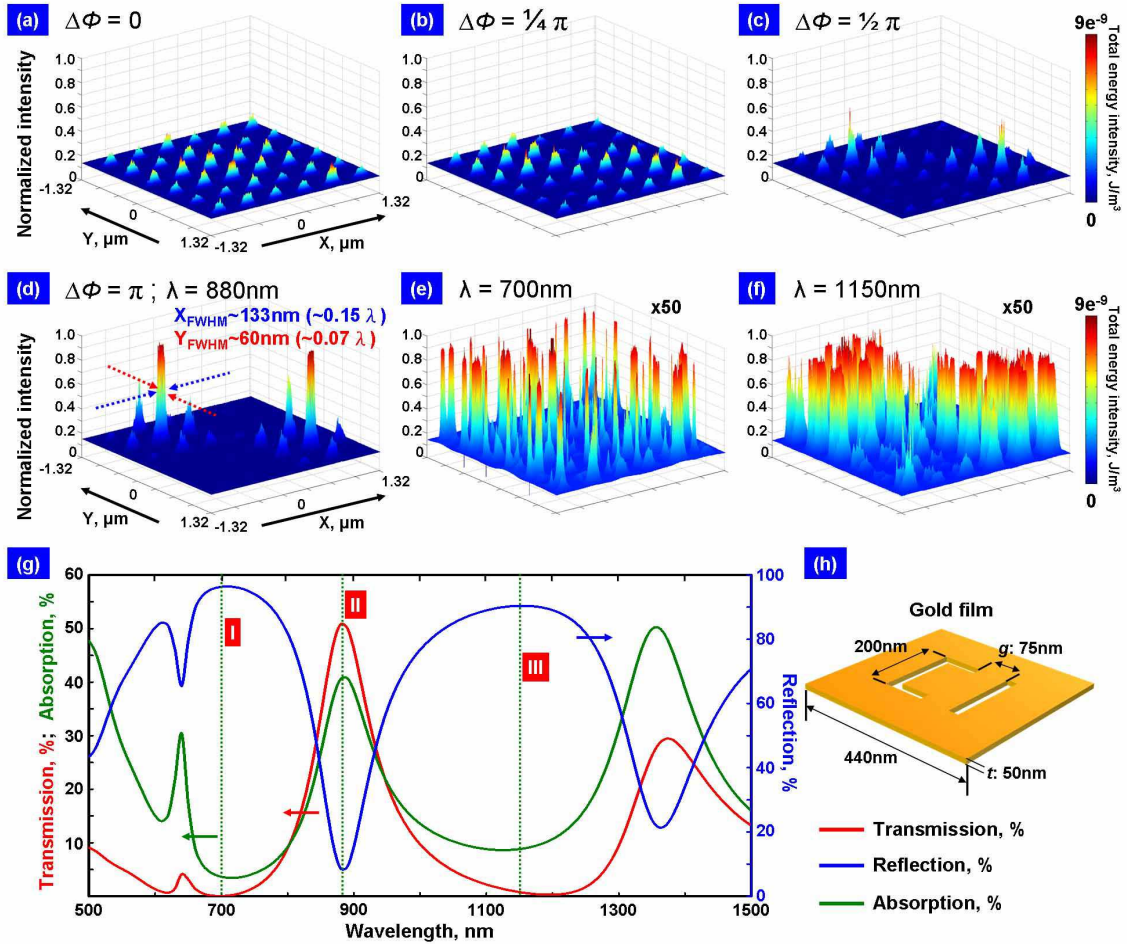}
  \caption{Coherent control of energy localization at metamaterial array of interacting plasmonic resonators. Colour maps (a)-(d) show the build-up of light localization with the increase of sinusoidal phase modulation of the driving field, $\Delta\phi$. Note two intensity peaks formed at the opposite quadrants of the landscape at resonance conditions at $\lambda$ \unit{880}{\nano\metre} and $\Delta\phi = \pi$. Maps (d)-(f) show the light localization effect at different wavelength for $\Delta\phi = \pi$: tuning the excitation wavelength away from the resonance (e) $\lambda$ = \unit{700}{\nano\metre} and (f) $\lambda$ = \unit{1150}{\nano\metre} destroys the localization. Figure (g) show the far-field transmission, reflection and absorbtion spectra of the metamaterial array and its unit cell (h). Localization is most pronounced at $\lambda$ = \unit{880}{\nano\metre} corresponding to the absorbtion resonance $II$.}
  \label{localization}
\end{figure*}

We demonstrate that controllable nanoscale localization of optical energy can be achieved at an artificial nanostructured  metamaterial surface, consisting of an array of altimetrically split ring plasmonic resonators, when such structure is illuminated with a coherent continuous light wave (a laser) those spatial variation of the phase at the metamaterial plane can be tailored. The required spatial variation of phase are slow and shall take place on the scale longer than the wavelength of light. This can be routinely achieved experimentally using widely commercially available liquid crystal spatial light modulators. The choice of asymmetrically split ring array as the nanoscale landscape was motivated by the recent studies that established a key role of inter-meta-molecular interaction in forming the electromagnetic response of such metamaterials \cite{fedotov2010,
papasimakis2009}. We will see below that the existence of strong interactions between plasmonic resonators is the key requirement for nanoscale localization. Figure~\ref{conf} shows artistic representation of the metamaterial array excited by the wave with a tailored phase profile creating a hot-spot at the other side of the array, at its immediate proximity.

Our analysis of this system is based on direct quantitative numerical simulation using a 3D Maxwell solver illustrated by qualitative modeling the metamaterial as a regular array of discrete interacting dipolar scatterers. In all numerical simulations we concentrated on the visible and the
near-field parts of the spectrum, assumed that metamaterial structures are constructed of gold plasmonic resonators and used realistic
material parameters and Joule loss factors by using the well established data for the dielectric parameters of the metal \cite{johnson1972}.

We demonstrate the idea of coherent control of light localization using the excitation beam with a periodic phase modulation across its
wavefront: even this most simple profile leads to a profound light localization at the array of interacting meta-molecules at certain
frequencies of the driving field. Fig.~\ref{localization} shows intensity maps at the metamaterial array driven by a monochromatic field manipulated such that, upon arrival at the metamaterial surface, the electric field has constant amplitude, a fixed polarization along $y$-direction, and a spatially varying phase $\phi(x,y) = (\Delta\phi_{max}/2)\sin(\kappa x) \sin(\kappa y)$, where $\kappa = 2\pi/(6 a)$ and $a$ is the spacing between adjacent meta-molecules. This corresponds to the incident wavefront modulated along $x$ and $y$ direction with a period of six unit cells of the metamaterial array. The array is a gold film of \unit{50}{\nano\metre} thickness perforated with \unit{25}{\nano\metre} slits of split square shape, as presented in Fig.~\ref{localization}(h). The metamaterial unit cell is $\unit{200} {\nano\metre}\times \unit{200} {\nano\metre}$.  The metamaterial nanostructure has a well defined plasmonic absorption resonance at \unit{880}{\nano\metre} corresponding to a local maximum in transmission and a minimum in reflection \cite{khardikov2010}.
We investigated field localization in the immediate proximity to the array at this wavelength of excitation and at two other wavelengths above (\unit{1150}{\nano\metre}) and below (\unit{700}{\nano\metre}) the resonance, at local minima of absorption and transmission.

\begin{figure}
\linespread{1}
\includegraphics[width=80mm]{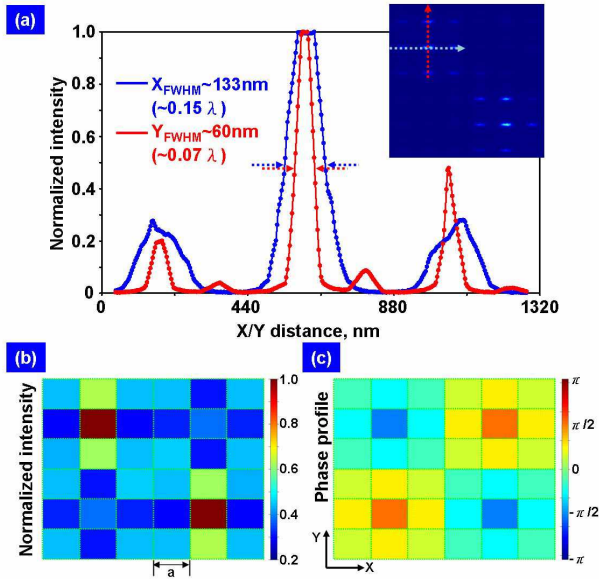}
\caption{The main cross-sections of the energy hot-spot (a) created at the metamaterial landscape (see Fig.2d) indicate a sub-wavelength localization with the total footprint at the peak of about 1 percent of $\lambda^2$. Colour maps (b) and (c) show the intensity and phase of excitations within the model of individual interacting dipoles representing the meta-molecules of array.}\label{resonant}
\end{figure}

The main characteristic features of the coherent control process are illustrated on Fig.~\ref{localization}. Fig.~\ref{localization}(b-c-d-e) show the intensity distribution at the array as a function of the amplitudes of spatial phase modulation. As expected, for a plane wave front of the incident beam ($\Delta\phi_{max}=0$) all meta-molecules in the array are excited equally. With an increase of the modulation depth, energy tends to concentrate at two opposite quadrants of the ${6}\times{6}$ section of the array. At ($\Delta\phi_{max}=\pi$) two hot-spots are formed which are
localized at a single meta-molecule six unit cells apart from a neighboring hot-spot along both $x$ and $y$ directions. For a fixed amplitude of phase modulation the localization is most pronounced at the plasmonic absorption resonance at \unit{880}{\nano\metre}: de-tuning away from the resonance destroys the localization, as illustrated on Fig.~\ref{localization}(e-f-g). Fig.~\ref{resonant}(a) shows intensity cross-section of the resonant hot-spot (\unit{880}{\nano\metre}) which has a footprint (at half maximum) of only $\unit{60}{\nano\metre}\times \unit{133}{\nano\metre}$
($\unit{0.0078}{\micro\metre}^2$).

The main features of the coherent control may be illustrated by a simple model in which we treat current oscillations in individual meta-molecules as point-like dipoles. An incident driving field of a given wavelength excites the current oscillations, producing scattered radiation.
These emitted fields in turn drive the corresponding oscillations in neighboring meta-molecules. The system dynamics can be described by the monochromatic version of Maxwell's wave equations, for instance,
\begin{equation}
(\nabla^2+k^2) \spvec{D}^+=-\nabla\times(\nabla\times\spvec{P}^+)
-ik\nabla\times \spvec{M}^+
\end{equation}
where $\spvec{D}^+(\spvec{r},t)$, $\spvec{P}^+(\spvec{r},t)$ and $\spvec{M}^+(\spvec{r},t)$ are the positive frequency components of the electric displacement, polarization, and magnetization densities, respectively. For an array meta-molecules labelled by an index $j$, at position $\spvec{r}_j$, with electric dipole moments $\spvec{d}_j(t)$ and magnetic dipole moments $\spvec{\mu}_j(t)$, the polarization and magnetization
densities are given by $\spvec{P}(\spvec{r},t) = \sum_j \spvec{d}_j(t) \delta (\spvec{r} - \spvec{r}_j)$ and $\spvec{M}(\spvec{r},t) = \sum_j
\spvec{\mu}_j(t) \delta (\spvec{r} - \spvec{r}_j)$, respectively. This model accounts for short range quasi-static dipole-dipole interactions between meta-molecules and long range interactions mediated by the radiated field and radiation losses. As a result, the material displays collective modes of oscillation, each with a distinct resonance frequency and collective radiative damping rate. The majority of collective modes do not couple directly to the uniform driving field because only collective modes with translational symmetry can couple to such a field. However, by allowing spatial variation in the driving field (for instance a spatial phase modulation, as illustrated above), one shall be able to excite a superposition of collective modes. A superposition of such modal excitations can lead to the formation of hot-spots located only at a few isolated meta-molecules.

The effect of strong inter-meta-molecular interactions in the energy localization may be illustrated by a simple example of two electric dipoles with a separation much less than the wavelength of light. The collective state of the dipoles exhibits super-radiant (enhanced radiative damping rate)
and sub-radiant (suppressed radiative damping rate) eigenmodes. In the first eigenmode the two dipoles are in-phase and in the second one they are $\pi$ out-of-phase. If we could prepare an incident field that excites an equal superposition of the two modes, it could then localize the energy in one of the dipoles. The crucial part in the emergence of the sub-radiant and super-radiant states in a two-dipole system are {\em recurrent} scattering events \cite{Lagendijk,RuostekoskiJavanainenPRA1997L} -- processes in which a wave is scattered more than once by the same dipole -- which cannot be described by the standard continuous medium electrodynamics, necessitating a discrete scatterer model of the metamaterial system. We present a less trivial example in the supplementary material. A numerical example of field localization achieved within the dipole model and showing
all characteristic features of the coherent process control is illustrated in Fig.~\ref{resonant}(b).

The opportunity to localize energy hot-spot at a single isolated meta-molecule provides for interesting applications. Indeed, by simply shifting phase profile function of the driving beam $\phi(x,y) = (\Delta\phi_{max}/2)\sin(\kappa x +\delta x) \sin(\kappa y+\delta y)$ on $\delta x$ and $\delta y$, integers of the lattice period, one can move the hot-spot at will across the metamaterial landscape. In what has been illustrated above, the hot-spots can be spaced by at least six unit cells from one another and thus moved at will in this space.
Our computational capabilities do not allow for numerical modeling of driving fields modulated at a larger scale which would involve modeling of a larger array of meta-molecule. We argue, however, that similar localization effects could take place in larger arrays where individual hot-spots are localized in much larger dim areas. Moreover, we argue that localization will be improved by appropriate selection of metamaterial and by controlling losses (for instance by using silver instead of gold) that will ensure long-range interactions between meta-molecules.

Controlling the hot-spot position on the metamaterial chess-board may be used for high-density data storage application in particular with phase-change nano-particles \cite{soares2007} placed at the meta-molecules and switched at the hot-spot. Moreover, the light localization technique can be used in a new type of optical imaging instruments with sub-wavelength resolution: one can envisage that an object of study (say, a cell) is placed on the metamaterial surface with a light-collecting optics and a photo detector above it. Moving the hot-spot from one place to another shall lead to the hot-spot being exposed to different scattering/absorption regimes which will affect intensity at the photo detector. A map of such intensities can be build to reconstruct a sub-wavelength image of the object. Here the metamaterials lattice pitch defines imaging resolution.

In conclusion, we demonstrated coherent control of light localization in photonic metamaterial. Well isolated energy hot-spots as small as $\lambda/10$ can be created and positioned at will on the plasmonic landscape by illuminating an array of interacting plasmonic resonators with a coherent continuous light beam with a tailored spatial variation of the phase profile.

The authors are grateful to N. Papasimakis, E. Rogers and S. Savo for numerous discussion and would like to acknowledge the financial support of the Engineering and Physical Sciences Research Council, UK and the Royal Society.

\bibliographystyle{nature}
\bibliography{metamaterials}

\clearpage

\textbf{\large Supplementary online material}\\

Here we provide a theoretical description of the model used to describe collective excitations in an ensemble of identical meta-molecules.

\appendix

\section{Theoretical Model}
\label{sec:theMod}

We consider an ensemble of $N$ meta-molecules placed at positions $\spvec{r}_j$ (~$j=1,\ldots,N$~) arranged in a two-dimensional rectangular lattice. This lattice is driven by an external, phase modulated, beam with electric field $\spvec{E}_{in}(\spvec{r},t)$ as described in the text. In general, the charge and current densities for the single-molecule oscillatory modes are described by a polarization density $\spvec{P}_j(\spvec{r},t)$ and magnetization density $\spvec{M}_j(\spvec{r},t)$. For simplicity, we assume that the spatial extent of a meta-molecule is much smaller than the wavelength of light and that the electric quadrapole moment of the meta-molecule vanishes. Each meta-molecule is described by a single mode of current oscillation whose contribution to the scattering of electromagnetic wave we then approximate by its electric and magnetic dipole moments. The state of the meta-molecule $j$ is represented by the dynamic variable $Q_j(t)$ with units of charge and whose spatial characteristics have been separated out into time independent functions $\spvec{\psi}(\spvec{r})$ and $\spvec{\varpi}(\spvec{r})$. Explicitly,
\begin{align}
  \spvec{P}_j &= Q_j(t) \spvec{\psi}(\spvec{r}-\spvec{r}_j)\simeq
  \spvec{d}_j \delta(\spvec{r}-\spvec{r}_j)\\
  \spvec{M}_j &= I_j(t) \spvec{\varpi}(\spvec{r}-\spvec{r}_j)\simeq
  \spvec{\mu}_j \delta(\spvec{r}-\spvec{r}_j)
\end{align}
where $I_j(t) = dQ_j/dt$ is the current. In this work, we assume the electric dipoles are aligned with the incident electric field and that the magnetic dipoles are oriented along the driving field's propagation direction so that only the electric dipoles are driven by the incident beam.

The coupled dynamic equations for the current oscillations and
electromagnetic fields can be systematically derived from the Lagrangian
formalism and will be presented elsewhere\cite{JenkinsLongPRB}.
In the absence of radiation losses, each current oscillation behaves as
a simple LC circuit with self capacitance $C$ and self inductance $L$.
The emitted radiation, results in a damping of the current
oscillation, which we treat here using the phenomenological
line-widths $\Gamma_E$ and $\Gamma_M$ associated with electric and
magnetic dipole radiation respectively.
In the multi-molecule system, however, the radiated fields couple to the dipole
moments of all other molecules in the system.
It is convenient to represent the current oscillations in terms  the dimensionless normal variables
\begin{equation}
  \label{eq:normVar}
  b_j(t) = \frac{e^{i\omega_0 t}}{\sqrt{2\omega_0}} \left(\frac{Q_j}{\sqrt{C}}
  +i\frac{\Phi_j}{\sqrt{L}}\right) ,
\end{equation}
where $\Phi_j=\int d^3 r\spvec{B}\cdot \spvec{\varpi}$ describes the
effective magnetic flux through the meta-molecule $j$, and $C$ and $L$
are the effective self capacitance and self inductance of a single
meta-molecule.
When the evolution of the input driving fields and the
meta-molecules themselves have frequency components in a narrow
bandwidth $\delta\Omega$ about the resonance frequency
$\omega_0=1/\sqrt{LC}$ such that $\delta\Omega/\omega_0 \ll 1$, we may
neglect any frequency dependence spontaneous emission rates or coupling
coefficients between the molecules.
Further, assuming that $\Gamma_E$, $\Gamma_M$, $\delta\Omega \ll
\omega_0$, we may make the rotating wave approximation, in which we
neglect terms oscillating at high frequencies $\sim 2\omega_0$.
In these limits, the equations of motion are
\begin{equation}
  \label{eq:2}
  \dot{\colvec{b}}(t) = \mathcal{C} \colvec{b}(t) + \colvec{F}(t) \text{,}
\end{equation}
where $\colvec{b} = (b_1,\ldots,b_N)^T$,  the coupling matrix
\begin{align}
  \mathcal{C} =& -\frac{\Gamma_E + \Gamma_M}{2}\mathbf{1} +i\frac{3}{4}
  \left(\Gamma_E \mathcal{G}_E + \Gamma_M\mathcal{G}_M \right)
  \nonumber\\
  & +
  \frac{3}{4} \sqrt{\Gamma_E \Gamma_M}\left(\mathcal{G}_\times +
    \mathcal{G}_\times^T\right) \text{,}
    \label{eq:CopulingMatC}
\end{align}
the driving field contribution is given by
\begin{equation}
  \label{eq:ForcingFunc}
  \colvec{F} = i\frac{e^{i\omega_0 t}}{\sqrt{2\omega_0}}
  \frac{\mathcal{E}_{in}(t)}{\sqrt{L}} \text{,}
\end{equation}
and $\colvec{\mathcal{E}}_{in}$ is the electromotive force
induced by the driving field  $\mathcal{E}_j =
\int d^3r\, \spvec{E}_{in}(\spvec{r},\Omega) \cdot
\spvec{\psi}(\spvec{r}-\spvec{r}_j)$,
The dimensionless coupling matrices $\mathcal{G}_E$, and $\mathcal{G}_M$ result from interactions with the  electric or
magnetic fields scattered from the electric or magnetic dipoles respectively, while the matrix $\mathcal{G}_\times$ accounts for the electric (magnetic) fields produced by the magnetic (electric) dipoles. Equation (\ref{eq:2}) corresponds to the integral representation of Maxwell's wave equations, as described by Eq.(1) in the text, and can be efficiently solved as a linear system. In the limits we've considered, there exist as many distinct collective eigenmodes of oscillation as there are meta-molecules in the system. Each collective mode corresponds to an eigenvector of the matrix $\mathcal{C}$ and has a distinct resonance frequency and decay rate, determined by the imaginary and real parts respectively of the corresponding eigenvalue. In the context of this simplified model, the scheme we proposed in
this manuscript boils down to choosing a spatial profile and detuning of the driving field so as to excite a linear combination of these eigenmodes such that the steady state solution is localized on the meta-molecular crystal.

\section{Effect of collective modes}

\begin{figure}
\includegraphics[width=0.8\columnwidth]{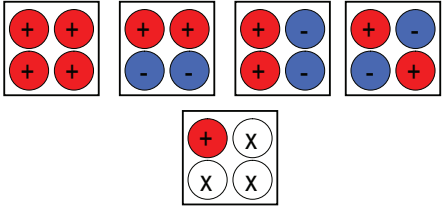}\vspace{-0.2cm}
\caption{First row: The four eigenstates of a $2\times2$ metamaterial system $\phi_j$ ($j=1,\ldots,4$), respectively.
The $+$ ($-$) sign refers to the positive (negative) amplitude of the current oscillation mode.
The second row: a superposition state $\Phi=\sum_j {1\over 2} \phi_j$ dominantly exciting one
of the dipoles. Here the sites marked by x are not excited.
} \label{dipolefig}
\end{figure}

The effect of strong interactions between different meta-molecules and the resulting collective radiative modes can be illustrated by a simple example of a $2\times2$ metamaterial array. If the inter-dipole separation is much smaller than the wavelength of light, the resonance frequencies and the radiative damping rates are very different from the ones of
an isolated dipole and they are affected by recurrent scattering processes in which the wave is scattered several times by the same dipole. The four energy eigenmodes of the system can be represented as shown in Fig.~\ref{dipolefig}.
Here the positive (negative) amplitude of the current oscillation mode is displayed as $+$ ($-$) sign.

If an incident field can be prepared in such a way that it excites an equal superposition of the four eigenmodes $\Phi=\sum_j {1\over2}\phi_j$, then one can immediately observe the approximate cancelation of the optical excitation energy in three of the four dipoles in the $2\times2$ metamaterial array. This simple example illustrates the localization in one of the dipoles.


\end{document}